\documentclass{epl}

\title{Dynamics of quantum trajectories in chaotic systems}

\author{D. A. Wisniacki\inst{1,2} \and F. Borondo\inst{1} \and
   R. M. Benito\inst{3}}

\institute{
  \inst{1} Departamento de Qu\'{\i}mica C--IX,
           Universidad Aut\'onoma de Madrid,
           Cantoblanco, 28049--Madrid (Spain). \\
  \inst{2} Departamento de F\'\i sica ``J.J. Giambiagi'',
           FCEN, UBA, Pabell\'on 1, Ciudad Universitaria,
           1428 Buenos Aires, Argentina. \\
  \inst{3} Departamento de F\'{\i}sica,
           E.T.S.I.\ Agr\'onomos, Universidad Polit\'ecnica de Madrid,
           28040 Madrid (Spain).
}

\begin{document}

\maketitle

\begin{abstract}
Quantum trajectories defined in the de Broglie--Bohm theory
provide a causal way to interpret physical phenomena.
In this Letter, we use this formalism to analyze the short time
dynamics induced by unstable periodic orbits in a classically
chaotic system, a situation in which scars are known to play a
very important role.
We find that the topologies of the quantum orbits are much more
complicated than that of the scarring and associated periodic orbits,
since the former have quantum interference built in. Thus  scar
wave functions are necessary to analyze the corresponding dynamics.
Moreover, these topologies imply different return routes to the
vicinity of the initial positions, and this reflects in the
existence of different contributions in each peak of the survival
probability function.
\end{abstract}

Trying to overcome interpretative difficulties \cite{Jammer}
encountered in the standard quantum theory \cite{vonNeuman},
Bohm developed in the 1950's an alternative formalism,
based on the notion of quantum trajectories \cite{Bohm1}.
This complementary quantum theory of motion \cite{Holland} combines
the accuracy of the standard quantum description with the intuitive
insight derived from the causal trajectory formalism,
thus providing a powerful tool to understand the physical mechanisms
underlying microscopic phenomena.
A key property of this theory is non--locality, since particles are
guided along their paths by a pilot wave, constituted by the system
wave function \cite{Broglie}.

The above mentioned characteristics are the cause of the vigorous
revival that the method is presently experiencing.
Taken from its computational side, it has been succesfully applied to
treat different realistic physical systems \cite{Askar,beswick,nos,ssr}.
This way of proceeding allowed, for example, Wyatt \cite{Wyatt}
to analyze tunneling of a wave packet through a barrier,
nicely elucidating the corresponding mechanism in terms
of quantum trajectories.
A portion of them, in the foremost part of the initial packet,
overcome the obstacle because they acquire some extra kinetic energy
during a boost phase, shortly after the launching of the particle.
With a similar analysis in mixed quantum--classical simulations,
Prezhdo and Brooksby \cite{Prezhdo} solved the quantum backreaction
problem for the dissociation reaction of O$_2$ on Pt.

Since the de Broglie--Bohm (BB) theory considers trajectories in a
fully quantum framewok, another line of research to which it has been
naturally applied is quantum chaos \cite{gut,Haake}.
Using this idea, several authors \cite{BBchaos} have followed the
separation of nearby Bohmian trajectories, thus computing quantum
analoges to Lyapunov exponents.
In our opinion, this way of proceeding is too simple.
The relative behavior of two particular trajectories do not mean,
in general, anything in view of the hollistic character \cite{Holland}
imposed by the non--local character of BB theory \cite{ssr}.
Moreover, the quantum Lyapunov exponents so defined are quantities
introduced in the context of infinitely long times, while it is widely
recognized that many interesting quantum phenomena, such as scars
\cite{Heller1}, have its origin in the short time dynamical evolution
of wave packets, much before the so called log time is reached.
The relevant information about short time recurrences,
and other quantum time scales, is contained in the survival probability,
$S(t)=|\langle \phi(t) | \phi(0) \rangle|^2$.

In this Letter we present a study of the topology and characteristics
of quantum trajectories computed using the BB theory in a situation
in which the dynamics of the initial state is strongly determined
by a classically chaotic PO.
The idea is to take advantage of the intuitive character of the BB
mechanics in order to provide a good understanding
of the quantum short time dynamics taking place in the process,
complementary to that provided by the usual scar theory
\cite{gut,Heller1,Heller2}.

The fundamental equation in the BB theory is derived from Madelung
hydrodynamical formulation of quantum mechanics \cite{Madelung},
which recasts the wave function in polar form,
%
\begin{equation}
 \phi(\textbf{r},t) \! = \! R(\textbf{r},t) \;
  {\rm e}^{{\rm i} S(\textbf{r},t)}
 \label{eq:1}
\end{equation}
(throughout the paper $\hbar$ is set equal to 1).
Quantum trajectories are then defined by means of expression
%
\begin{equation}
   m \dot{\bf r} = \nabla S = {\rm Im} (\nabla \phi/\phi),
 \label{eq:qt}
\end{equation}
which can be numerically integrated once $\phi$ is known.
Alternatively, one can consider the corresponding newtonian form,
$m \ddot{\bf r} = -\nabla(V+Q)$, where the extra term,
$Q=-(1/2m)(\nabla^2 R/R)$ called quantum potential,
is responsible for all quantum effects taking place in the process.

\begin{figure}[tb]
 \onefigure[scale=0.5]{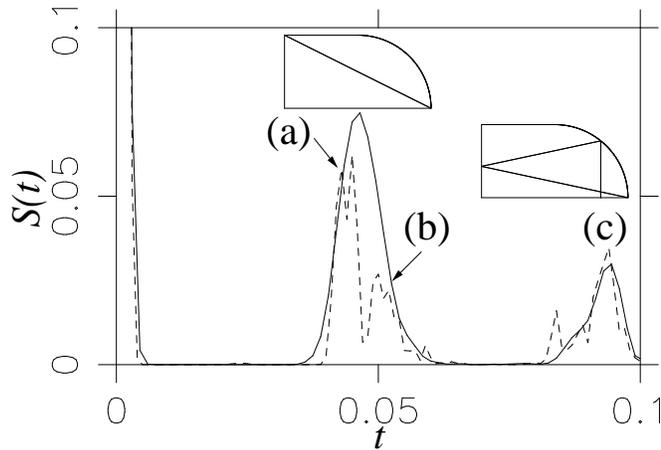}
 \caption{Survival probability function corresponding to a wave packet
  initially centered on the diagonal periodic orbit of a desymmetrized
  stadium billiard with $r=1$ and area $1+\pi/4$,
  calculated both exactly (full line), and with the quantum trajectories
  approximation defined in the text (dashed line).
  The insets show the two periodic orbits responsible for the two
  recurrence peaks observed.}
 \label{fig:1}
\end{figure}

The problem that we have chosen to study is the dynamics of a
particle of mass 1/2 enclosed in a desymmetrized stadium billiard of
radius $r=1$ and area $1+\pi/4$ with Dirichlet boundary conditions,
influenced by the diagonal periodic orbit (PO).
This corresponds to a diamond shape in the full version of the stadium.
For this purpose, we start from a harmonic oscillator coherent state
%
\begin{equation}
   \phi(0) = (2\alpha/\pi)^{1/4}
    {\rm e}^{-\alpha(x-x_0)^2 -\alpha(y-y_0)^2 +
    {\rm i}(P_x^0 x+P_y^0 y)},
 \label{eq:phi0}
\end{equation}
with $\alpha=30.68$, initially located on the middle of the diagonal,
$(x_0,y_0,P_x^0,P_y^0)$ = $(1,1/2,96/\sqrt{5},-48/\sqrt{5})$.
From these values we have $E=2304$ and a period $T=0.0466$.
As stated in the introduction, the associated short time dynamics can
be adequately discussed in terms of the survival probability, $S(t)$,
which is shown in full line in Fig.\ \ref{fig:1} up to $t=0.1$.
%
As can be seen, the curve consists of two peaks, taking place at
$t\sim 0.047$ and $\sim$0.094, respectively, originated by the first
two recurrences of the initial packet (\ref{eq:phi0}).
Also, the second peak shows some substructure, in the form of two
shoulders located to the left of the main maximum.
These features were explained \cite{Diego} in terms of the scarring
effect of the dynamics along the diagonal PO, using dynamically
averaged wave packets \cite{Polavieja}.
With this method, we were able to uncover the role played by other short,
but longer, POs dynamically related to the diagonal one.
In the inset to Fig.\ \ref{fig:1} we have represented the diagonal PO,
contributing to both recurrences in $S(t)$, plus the extra PO
contributing also to the second one.

In this Letter we want to go step further and provide new insight in
these results, using the interpretative abilities of the BB theory
to understand quantum effects \cite{Wyatt,Prezhdo}.
For this purpose, we will use an approximate version of the survival
probability, defined in the following way.
Starting from the phase space representation in terms of the Wigner
function, $W$, associated to $\phi$,
%
\begin{equation}
   S(t) = \int \int {\rm d} \textbf{q} {\rm d} \textbf{P} \;
          W_0(\textbf{q},\textbf{P}) W_t(\textbf{q},\textbf{P}),
 \label{eq:S(t)}
\end{equation}
$W$ is substituted by the following approximate distribution based
on quantum trajectories, [$x_i(t),y_i(t)$],
%
\begin{equation}
\begin{array}{rl}
 \rho_t (x,y,P_x,P_y) = N^{-1} \displaystyle \sum_i^N & \delta[x-x_i(t)] \; \delta[y-y_i(t)] \\
    & \times \displaystyle \delta \left[P_x - \left(\frac{\partial S}{\partial x}\right)_{x=x_i(t)} \right]
      \displaystyle \delta \left[P_y - \left(\frac{\partial S}{\partial y}\right)_{y=y_i(t)} \right],
\end{array}
 \label{eq:rho}
\end{equation}
Finally, and in order to make $\rho$ a smooth function, gaussian
distributions on the delta functions are allowed, thus obtaining
%
\begin{equation}
\begin{array}{rl}
   S(t) \approx N^{-1} \displaystyle \sum_{i,j}^N &
      \exp\{-\sigma [x_i(t)-x_j(0)]^2 - \sigma [y_i(t)-y_j(0)]^2  \\
      & -[P_{xi}(t)-P_{xj}(0)]^2/\sigma-[P_{yi}(t)-P_{yj}(0)]^2/\sigma\}.
\end{array}
 \label{eq:ssuma}
\end{equation}

The results obtained with this approximation are shown in dashed line
in Fig.\ \ref{fig:1}.
They have been computed using 80 quantum orbits, obtained by numerical
integration of eq.~(\ref{eq:qt}), for which a value of $\sigma=156.25$
is adequate.
Some caution must be exerted when integrating the quantum equations
of motion in cases like ours, for which the dynamics of the system
are very chaotic.
As will be seen below, the time--dependent wave packet develops very
quickly a tremendous complexity, showing a complicated nodal structure
of size of the order of $\hbar$.
In the present case we have chosen a Gear stiff integration method
with tolerance control.
The trajectories were initially distributed uniformly over four radii:
0.00333, 0.0333, 0.05 and 0.1 respectively, around $(x_0,y_0)$.
As can be seen in Fig.\ \ref{fig:1}, the agreement between the global
shape of the exact and approximate survival probabilities is good,
except for some discrepancies in the relative intensities.
Moreover, both calculations present recurrences at exactly the same times,
and the latter even reproduces the shoulder structure existing in
the exact second recurrence.
It should be stressed at this point that our approximate results can
be easily converged to the exact ones, by using a better (gaussian)
sampling procedure of the initial conditions.
A full account of this treatment will be given elsewhere \cite{Larita}.
However, this is not an essential point for the discussions of
this Letter, in which we prefer, for the sake of the discussions below,
to use our imperfect sampling, that artificially hide the importance of
quantum trajectories arriving to the first recurrence at average times,
in favour of those making it either too early or too late.
Due to this fact, our approximate results (see Fig.~\ref{fig:1})
clearly show the existence of two contributions, which nicely reproduce
both tails of the first recurrence.

The fact that we are basing our analysis in an approximate expression
may be considered at first sight as a drawback.
However, it is quite the contrary, since in the way in which it is
defined, our version of $S(t)$ comes given as a sum of positive
contributions, this having the important advantage of allowing the
localization of those trajectories contributing the most to $S(t)$.
This makes our method a superb tool to investigate, in an intuitive
and straighforward way, how the dynamics of the initial wave packet
takes place.

\begin{figure}
 \twofigures[scale=0.35]{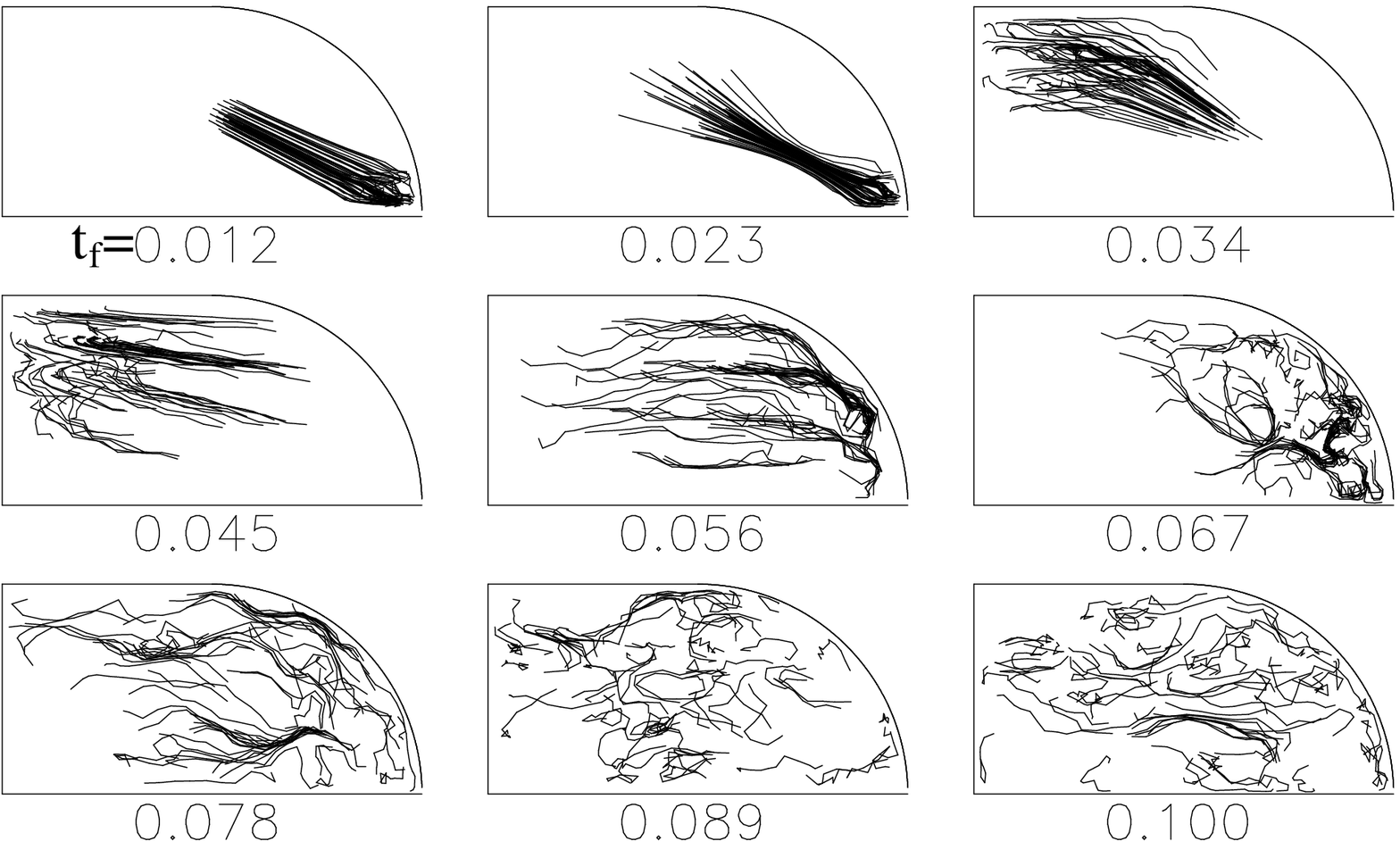}{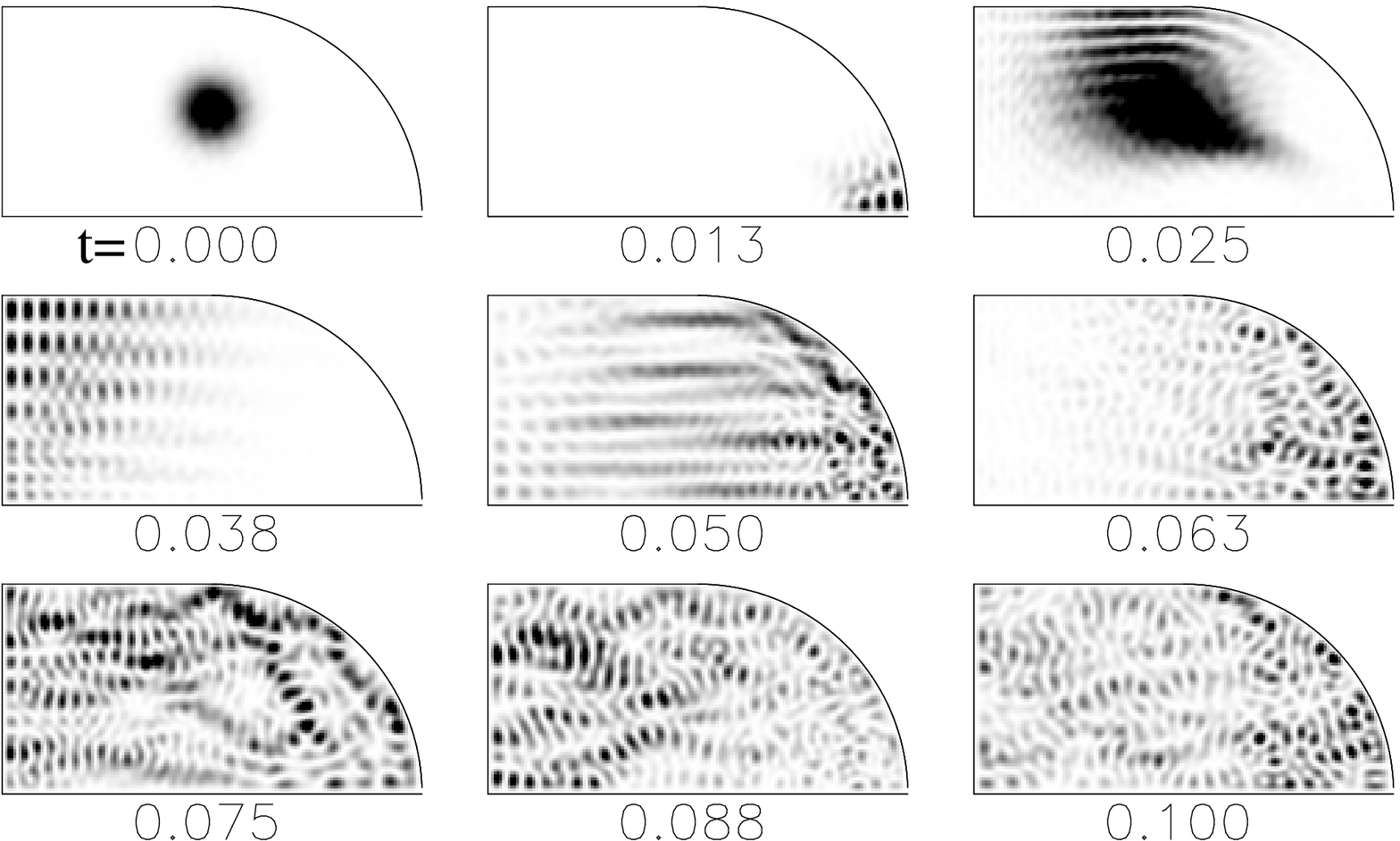}
 \caption{Propagation of 80 quantum trajectories corresponding to a
  wave packet initially centered on the diagonal periodic orbit of a
  desymmetrized stadium billiard with $r=1$ and area $1+\pi/4$.
  See text for details.}
 \label{fig:2}
 \caption{Time evolution of the probability density, $|\phi(x,y,t)|^2$,
  corresponding to the quantum trajectories of Fig.~\protect\ref{fig:2}.
  See text for details.}
 \label{fig:3}
\end{figure}

To discuss this in detail, let us start by examining the quantum
trajectories used in our calculation of $S(t)$, which are shown in
Fig.~\ref{fig:2}.
%
In order to make the figure clear, these trajectories have been
splitted into several consecutive segments, which are presented in
the different panels.
In this way, the first plot, labeled $t_f=0.012$, corresponds to paths
described by the trajectories during interval $t=0-0.012$, the second
in interval $t=0.012-0.023$, and so on.
To help in the discussion, it is interesting to consider also the time
evolution of the corresponding wave packet, which is shown in
Fig.\ \ref{fig:3}.
These results, that were used in the integration of eq.\ \ref{eq:qt},
have been obtained by projecting $\phi(0)$ on the stadium
eigenstates and subsequent application of the evolution operator.

As expected, trajectories (Fig.\ \ref{fig:2}) and wave packet
(Fig.\ \ref{fig:3}) follow, in the first part ot their evolution
($t \leq 0.023$), a very classical path, as dictated by Ehrenfest's
theorem.
After the first rebound at the lower right corner, they both spread
at $t\sim 0.025$ in a fan--type pattern, experiencing the well known
focalization effect originated by the self--focal point of the orbit
\cite{Heller2}.
Afterwards, the dispersed wave packet (see Fig.\ \ref{fig:3})
collides with the region around the upper left corner, giving rise to
a noticeable series of horizontal fringes, formed by the maxima and
nodal lines of the wave function.
These maxima originate flat plateaus in the associated quantum potential,
separated by deep canyons along the nodal lines.
This pattern greatly affect the topology of the corresponding quantum
trajectories, as can be seen in Fig.\ \ref{fig:2}.
Then, trajectories, during the interval $t \simeq 0.034-0.056$,
are mostly horizontal, following paths along the maxima of the
corresponding pilot wave (see central tier in Fig.~\ref{fig:3}).
Moreover, particles are ``kicked'' when trying to cross this structure,
giving rise to the short diagonal segments observed, for example,
in some trajectories at the top of panel labelled $t=0.034$.
Other rebounds take place at subsequent times, originating for
$t \geq 0.06$ a complicated structure in the distribution of the
quantum probability density.
The topology of the corresponding trajectories becomes accordingly
also very complex, showing an extreme sensitivity to both initial
conditions and the accuracy imposed to the integration procedure.
It should, however, be pointed out the extreme resemblance which is
obtained between the patterns formed by the maxima of $\phi(t)$ in
Fig.~\ref{fig:3} and the accumulations of quantum trajectories in
Fig.~\ref{fig:2}.
This is even more surprising when considering the small number of
trajectories used in our calculation.
The above mentioned dynamical sensitivity can be a problem when
trying to interpret the behavior of individual trajectories for
very long times, but it is reasonable to admit that the statistical
properties of ensembles of trajectories, such as that considered in
the approximate calculation of $S(t)$ [eq.\ (\ref{eq:ssuma})],
behave adequately, similarly to what happens in ergodic theory.

\begin{figure}
 \onefigure[scale=0.4]{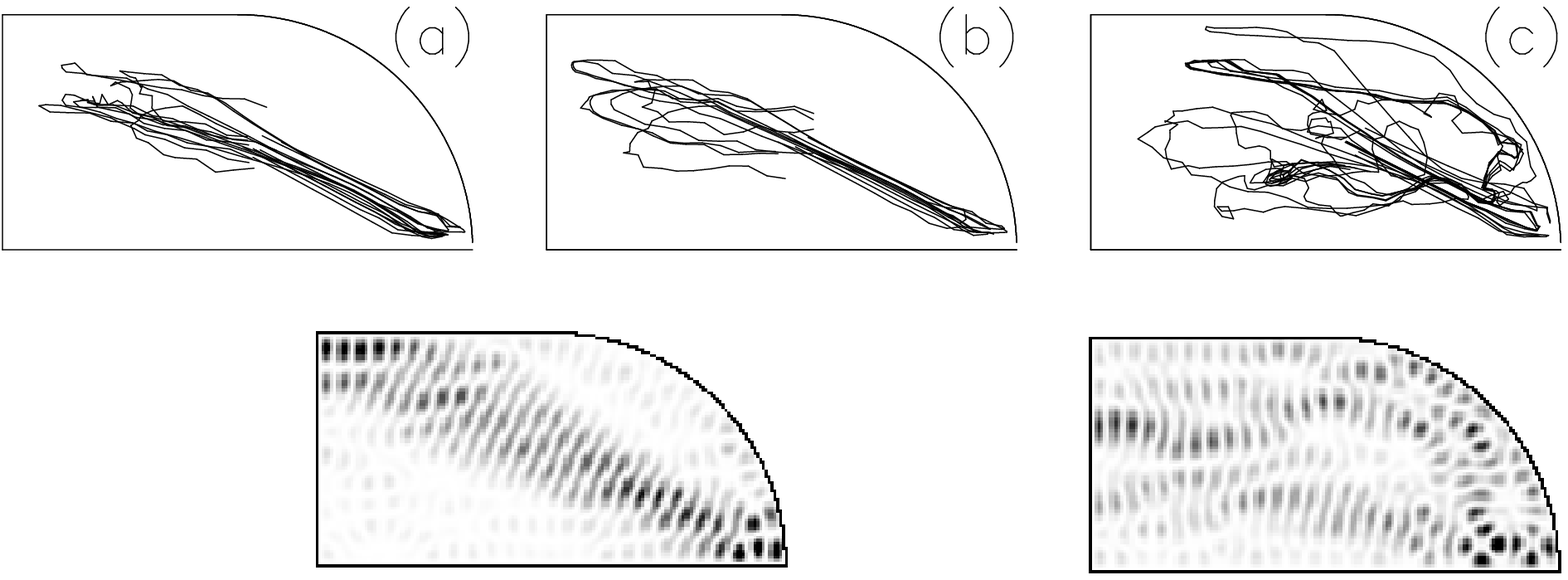}
 \caption{(Upper part) Quantum trajectories representing the largest
   contributions to the approximate version of $S(t)$
   [eq.\ \protect\ref{eq:ssuma}] for peaks (a), (b) and (c) in
   Fig.~\protect\ref{fig:1}. \protect \\
   (Lower part): Scar wave function, constructed with the average
   method of Ref.\ \protect\cite{Polavieja}, corresponding to
   trajectories in (a) and (b) (left part), and to (c) (right part).}
 \label{fig:4}
\end{figure}

After having discussed the general behavior of the quantum trajectories
used in our study, let us investigate now which one of them are more
relevant to explain the different features exhibited by the survival
probability.
Figure \ref{fig:4} shows such trajectories for the three labeled
peaks of $S(t)$ in Fig.~\ref{fig:1}.
%
Panels (a) and (b) corresponds to trajectories in the two separated
peaks at the first recurrence.
Those in (a) correspond to particles that arrive earlier to the
vicinity of the initial distribution, $\rho(0)$, thus corresponding
to what can be called forerunners.
As can be seen, the topology of the associated trajectories is
extremely similar to the classical diagonal PO, showing only a very
small dispersion, that mirrors the variation of the corresponding
initial conditions.

On the other hand, trajectories in panel (b), arriving later to the
initial region, trace slightly different paths.
At the upper left corner, they make a left turn that delays their
arrival to the final destination.
By doing this, they allow the pilot wave to spread more profusely,
also giving it enough time to develop the horizontal series of nodal
lines that are observed in Fig.~\ref{fig:3}.
As explained before, the existence of these nodal lines greatly affects
the paths taken by quantum trajectories, being the origin of the
ladder--type structure observed in the last part of trajectories
shown in Fig.~\ref{fig:4}(b).
This important quantum effect has never been discussed in the literature
before, and shows that the peaks in $S(t)$ consist of different
contributions with somewhat heterogeneous histories, that determine
the time delay with which they arrive to the initial region.
This is a nice example of how Bohmian mechanics can deal with the time
variable in a straighforward way \cite{time}, in contrast to what
happens in the standard quantum theory.
Also, these contributions can be traced back to their original launching
positions in the initial wavepacket, thus establishing a true causal
connection between initial and final conditions \cite{Larita}.

Let us analyze next the second recurrence at $t \sim 0.09$.
Now the situation is much more complicated, since as described in
Ref.~\cite{Diego} at least two other POs need to be considered.
In view of our approximate results, the most interesting feature to
considered here is the maximum, leaving aside the shoulder structure.
Figure \ref{fig:4}(c) shows the quantum trajectories contributing
the most to that part of $S(t)$.
As can be seen, an attempt to give an interpretation in the same
terms used before for the first recurrence is less convincing,
since the existence of a connection between quantum trajectories
and lines of maxima at the corresponding time (see panels in
Fig.~\ref{fig:3}) is not so obvious.
This is not unexpected since, as stated before, the dynamics here
get diluted into too many POs.

An alternative way to interpret these results is to use scar functions.
This is going one step further the classical dynamics of the POs,
which only represents a first approximation to the problem.
In quantum (or semiclassical) mechanics one has to include (at least
to some extend) the associated wave properties.
Scar wavefunctions can be constructed in a number of ways.
We will use the method of dynamical averaging introduced by us
\cite{Polavieja,Diego2}, that is very efficient.
A very elegant way to construct analogous functions has been discussed
by Vergini and Carlo in their resonance treatment of quantum chaos
\cite{Carlo}.
The scar wave function relevant for the discussion of the first
recurrence in $S(t)$ is presented in the left part of Fig.\ \ref{fig:4},
below the corresponding trajectories [panels (a) and (b)].
As can be seen, it presents the right physical characteristics to
explain the (slightly) different behavior of the two peaks uncovered
by our approximate calculation of the survival probability.
Namely, the function is on the one hand localized along the diagonal PO,
with a series of nodal lines perpendicular to it,
and a distribution of the probability density in agreement with the
expected focalization effect.
But more importantly, this function presents at the upper left corner
the series of horizontal fringes explaining the delayed behavior of the
trajectories contributing to the second peak of the first recurrence
in $S(t)$ [trajectories in (b)].

Finally, with respect to the maximum of the second recurrence
the corresponding scar wave function is shown in right part of
Fig.\ \ref{fig:4}.
In terms of it, the interpretation of trajectories in (c) is much
more obvious, and analogous to that given before.
The relevant quantum trajectories clearly follow the lines of maxima
of the associated scar wave function.
The reason for the agreement is clear.
The construction procedure of the scar function only includes dynamical
information concerning an isolated PO and its associated manifolds
\cite{Diego2}, thus making possible the comparison that we were
looking for.

In summary, we have presented in this Letter a first study on
the dynamical behavior of quantum trajectories, as defined in the
de Broglie--Bohm theory, for a system in which the classical
dynamics are highly chaotic.
Searching for the elusive concept of quantum chaos, this situation had
been previously studied by other authors in the infinite time limit.
Contrary, we have considered here the other limit, i.e.\  short time
dynamics, for which scarring effects are known to be important.
Proceeding in this way and taking advantage of the causal character
of our approach we have found new interesting features in the
corresponding survival probability.
In particular, we have shown how the peaks appearing in this function
at the recurrence times can be decomposed in different contributions
with heterogenous dynamical histories.

%

\acknowledgments

This work was supported by MCyT and MECD (Spain) under contracts
BMF2000--437 and SB2000--0340.
DAW greatfully acknowledges support from Fundaci\'on Antorchas
(Argentina).


\begin{thebibliography}{0}

  \bibitem{Jammer}
  \Name{M. Jammer}
  \Book{The Conceptual Development of Quantum Mechanics}
  \Publ{American Institute of Physics, Madison}
  \Year{1989}

  \bibitem{vonNeuman}
  \Name{J. von Neumann}
  \Book{Mathematical Foundations of Quantum Mechanics}
  \Publ{Princeton University Press, Princeton}
  \Year{1955}

  \bibitem{Bohm1}
  \Name{D. Bohm}
  \REVIEW{Phys. Rev.}{85}{1952}{166};
  \SAME{85}{1952}{194}.

  \bibitem{Holland}
  \Name{P. R. Holland}
  \Book{The Quantum Theory of Motion}
  \Publ{Cambridge University Press, Cambridge}
  \Year{1993}

  \bibitem{Broglie}
  \Name{L. de Broglie}
  \REVIEW{Ann. Phys. Paris}{3}{1925}{22}.

  \bibitem{Askar}
  \Name{F. Sales Mayor, A. Askar and H. A. Rabitz}
  \REVIEW{J. Chem. Phys.}{111}{1999}{2423}.

  \bibitem{nos}
  \Name{A. S. Sanz, F. Borondo, and S. Miret--Art\'es}
  \REVIEW{Phys. Rev. B}{61}{2000}{7743};
  \Name{A. S. Sanz, F. Borondo, and S. Miret--Art\'es}
  \REVIEW{Europhys. Lett.}{55}{2001}{303};
  \Name{A. S. Sanz, F. Borondo, and S. Miret--Art\'es}
  \REVIEW{J. Phys.: Condens. Matt.}{14}{2002}{6109}.

  \bibitem{beswick}
  \Name{E. Gindensperger, C. Meier, and J. A. Beswick}
  \REVIEW{J. Chem. Phys.}{116}{2002}{8};
  \Name{E. Gindensperger, C. Meier, J. A. Beswick, and M--C. Heitz}
  \REVIEW{J. Chem. Phys.}{116}{2002}{10051}.

  \bibitem{ssr}
  \Name{R. Guantes, R. Margalef--Roig, A. S. Sanz, F. Borondo,
  and S. Miret--Art\'es}
  \REVIEW{Suf. Sci. Rep.}{}{in press}{}.

  \bibitem{Wyatt}
  \Name{ R. E. Wyatt}
  \REVIEW{Phys. Rev. Lett.}{86}{2001}{3215}.

  \bibitem{Prezhdo}
  \Name{O. V. Prezhdo and C. Brooksby}
  \REVIEW{Phys. Rev. Lett.}{86}{2001}{32151}.

  \bibitem{gut}
  \Name{M. C. Gutzwiller}
  \Book{Chaos in Classical and Quantum Mechanics}
  \Publ{Springer--Verlag, New York}
  \Year{1990}

  \bibitem{Haake}
  \Name{F. Haake}
  \Book{Quantum Signatures of Chaos}
  \Publ{Springer--Verlag, Berlin}
  \Year{2001}

  \bibitem{BBchaos}
  \Name{R. H. Parmenter and R. W. Valentine}
  \REVIEW{Phys. Lett. A}{201}{1995}{1};
  \Name{S. Konkel and A. J. Malowski}
  \REVIEW{Phys. Lett. A}{238}{1998}{95}.

  \bibitem{Heller1}
  \Name{E. J. Heller}
  \REVIEW{Phys. Rev. Lett.}{53}{1984}{1515}.

  \bibitem{Heller2}
  \Name{E. J. Heller}
  \Book{Chaos and Quantum Physics}
  \Editor{M. J. Giannoni, A. Voros, and J. Zinn--Justin}
  \Publ{Elsevier, Amsterdam}
  \Year{1991}

  \bibitem{Madelung}
  \Name{E. Madelung}
  \REVIEW{Z. Phys.}{40}{1926}{332}.

  \bibitem{Diego}
  \Name{D. A. Wisniacki, F. Borondo, E. Vergini, and R. M. Benito}
  \REVIEW{Phys. Rev. E}{62}{2000}{R7583}.

  \bibitem{Polavieja}
  \Name{G. G. de Polavieja, F. Borondo and R. M. Benito}
  \REVIEW{Phys. Rev. Lett.}{73}{1994}{1613}.

%
  \bibitem{Larita}
  \Name{L. Hernando, D. Wisniacki, F. Borondo, and R. M. Benito}
  \REVIEW{}{}{in preparation}.

  \bibitem{time}
  \Editor{J. G. Muga, R. Sala and I. L. Egusquiza}
  \Book{Time in Quantum Mechanics}
  \Publ{Springer--Verlag, Berlin}
  \Year{2002}

  \bibitem{Diego2}
  \Name{D. A. Wisniacki, F. Borondo, E. Vergini, and R. M. Benito}
  \REVIEW{Phys. Rev. E}{63}{2001}{066220}.

  \bibitem{Carlo}
  \Name{E. G. Vergini and G. G. Carlo}
  \REVIEW{J. Phys. A}{34}{2001}{4525}.
%
\end{thebibliography}
\end{document}